\RequirePackage{fix-cm}
\documentclass[twocolumn,epjc3]{svjour3}  
\smartqed  
\RequirePackage{graphicx}
\RequirePackage{mathptmx}      
\RequirePackage{latexsym}
\RequirePackage{slashed}
\RequirePackage[numbers,sort&compress]{natbib}
\RequirePackage[colorlinks,citecolor=blue,urlcolor=blue,linkcolor=blue]{hyperref}
\usepackage{amsmath}
\usepackage{amsfonts}
\usepackage{amssymb}
\usepackage{url}
\usepackage{cancel}
\usepackage{comment}
\RequirePackage[caption=false]{subfig}
\RequirePackage{subfig}
\RequirePackage{blindtext, subfig}

\begin{document}

\title{Neutrino specific spin-3/2 dark matter
}


\author{Ashok Goyal\thanksref{e1,addr1}
        \and
        Mohammed Omer Khojali\thanksref{e2,addr2,addr3}
        \and 
        Mukesh Kumar\thanksref{e3,addr4}
        \and
        Alan S. Cornell\thanksref{e4,addr3}
}

\thankstext{e1}{e-mail: agoyal45@yahoo.com}
\thankstext{e2}{e-mail: khogali11@gmail.com}
\thankstext{e3}{e-mail: mukesh.kumar@cern.ch}
\thankstext{e4}{e-mail: acornell@uj.ac.za}


\institute{Department of Physics \& Astrophysics, University of Delhi, Delhi, India. \label{addr1}
           \and
           Department of Physics, University of Khartoum, PO Box 321, Khartoum 11115, Sudan.\label{addr2}
           \and
           Department of Physics, University of Johannesburg, PO Box 524, Auckland Park 2006, South Africa. \label{addr3}
           \and 
           School of Physics and Institute for Collider Particle Physics, University of the Witwatersrand,Johannesburg, Wits 2050, South Africa. \label{addr4}
}

\date{Received: date / Accepted: date}

\maketitle

\begin{abstract}
In this paper we consider a spin-$\frac{3}{2}$ dark matter (DM) particle which couples to neutrinos, as a viable candidate to produce the observed DM relic density through the thermal freeze-out mechanism. The couplings of DM to neutrinos is considered first in a general dimension-6 effective field theory framework. We then consider two specific neutrino-portal models discussed in the literature. In the first model DM couples to the standard model neutrinos through mixing generated by a sterile pseudo-Dirac massive neutrino, and in the second model we consider the $U(1)_{L_\mu - L_\tau}$ gauge symmetric model. For each of these models we explore the parameter space required to generate the observed relic density. The constraints on the parameters of these models from the existing and proposed neutrino experiments, as well as from existing cosmological and astrophysical bounds, are considered in the context of the relic density calculations.     
\end{abstract}

\section{Introduction}
\label{intro}
The twin problem of the existence of tiny neutrino masses necessitated by the observation of neutrino oscillations and the existence of dark matter (DM), making up roughly 75\% of the matter content of the Universe, presents a unique opportunity to seek a common framework of possible connections between these two sectors. Note that there have been several studies in the literature~\cite{Blennow:2019fhy,Batell:2017cmf,Bringmann:2013vra,Alvey:2019jzx,Boehm:2006mi,Baumholzer:2019twf,Patel:2019zky,Hagedorn:2018spx,Coito:2022kif} to explore the connection between DM particles which interact only with the Standard Model (SM) neutrinos through a mechanism for the generation of neutrino mass. The cosmological implications of such scenarios, the constraints, and scope from indirect searches have also been well discussed in Refs.~\cite{Boehm:2000gq,Boehm:2004th,Bertschinger:2006nq,Serra:2009uu,Wilkinson:2014ksa}.

The models in which DM phenomenology is dominated by its interactions with neutrinos do not suffer tensions from direct detection experimental constraints~\cite{Arcadi:2017kky}. However, it has been challenging to construct a model which not only accounts for tiny neutrino mass but at the same time couples to the SM neutrinos with a strength that produces efficient DM annihilation (as required for the observed DM relic density). The see-saw mechanism gives rise to small neutrino masses through the existence of sterile massive Majorana neutrinos. This generally results in small Yukawa couplings, unable to generate the requisite DM annihilation rate. In the absence of SM gauge interactions the DM still needs to have sizable interactions with the neutrinos. This leads to three possible SM portals to the dark sector {\it viz.} (i) the gauge-boson portal, (ii) the Higgs portal and (iii) the neutrino portal. In the neutrino portal a right-handed pseudo-Dirac sterile neutrino $N_R$ links the SM neutrinos with the dark sector. An approximate lepton-number symmetry is assumed to facilitate relatively large mixing with the SM neutrinos. The small neutrino-mass is achieved through small lepton number symmetry violation, as in a linear or inverse seesaw mechanism, as discussed in Refs.~\cite{Blennow:2019fhy,Batell:2017cmf}. Constraints on DM annihilation cross-sections into SM neutrinos from the existing and upcoming experiments have been discussed in Ref.~\cite{Arguelles:2019ouk}.

Recently there have been studies~\cite{Muong-2:2006rrc,Holst:2021lzm,Drees:2021rsg,Foldenauer:2018zrz,Biswas:2016yan} to link the measured anomalous magnetic moment of the muon by the Fermi-Lab collaborations~\cite{Muong-2:2021ojo} (which are found to be consistent with the earlier Brookhaven E821 collaboration results) with the study of DM phenomenology. These studies employ a $U(1)_{L_\mu-L_\tau}$ gauge symmetry under which the muon, as well as fermionic spin-$\frac{1}{2}$ DM, are assumed to carry equal $U(1)_{L_\mu-L_\tau}$ charge. The corresponding gauge-boson $Z^{\prime}_\mu$ acts as a mediator between the SM leptons ($\mu$ and $\tau$) and the DM. The DM direct detection bounds are evaded in these models. 
For the simultaneous resolution of the anomalous magnetic moment of the muon, and to obtain the required thermal relic density, the gauge boson mass was found to lie in the MeV range~\cite{Holst:2021lzm,Drees:2021rsg}.

Spin-$\frac{3}{2}$ particles have historically been one of the earliest DM candidates, when they appeared naturally in a local super-symmetric or super-gravity theory, such as gravitons. Later they were studied in the Rarita-Schwinger framework as massive spin-$\frac{3}{2}$ fields~\cite{Rarita:1941mf}. In recent years there have been several studies of spin-$\frac{3}{2}$ weakly interacting massive particles (WIMP) in a freeze-out DM scenario which contributes to the observed DM relic density in simplified models \cite{Yu:2011by,Ding:2013nvx,Khojali:2016pvu,Khojali:2017tuv}, in effective field theory (EFT) models and in a Higgs portal model~\cite{Christensen:2013aua,Ding:2012sm,Ding:2013nvx,Chang:2017gla,Chang:2017dvm}. Spin-$\frac{3}{2}$ particles interacting gravitationally through the mediation of spin-2 gravitons in the Randall-Sundram framework (as possible DM candidates) were considered in Ref.~\cite{Goyal:2019vsw}. Decaying spin-$\frac{3}{2}$ DM particles of mass $\sim 10^4$~GeV have been shown to yield the observed relic density during inflationary reheating through the scattering of SM particles~\cite{Garcia:2020hyo}. Spin-$\frac{3}{2}$ DM decaying into SM neutrinos and photons has been discussed in Ref.~\cite{Dutta:2015ega} in the context of the 3.55 keV galactic $X$-ray spectrum. 

In this paper we study the spin-$\frac{3}{2}$ dark matter in an EFT model and consider the most general dimension-6 effective Lagrangian relevant for DM annihilation into neutrinos. We then consider a simple neutrino portal model in which the DM couples to the SM neutrinos through mixing generated by a sterile pseudo-Dirac massive neutrino. The neutrino portal model is implemented in the $s$- and $t$-channels. Finally we consider a $U(1)_{L_\mu-L_\tau}$ gauge model as discussed above. For each of the three models we discuss the prospects of present and future neutrino detectors dedicated to probe the DM annihilation into neutrinos. See for example Ref.~\cite{Blennow:2019fhy} for details.

The paper is organised as follows: In sect.~\ref{framework} we summarize the spin-$\frac{3}{2}$ framework. An EFT Lagrangian with relevant dimension-6 operators for DM annihilation into neutrinos is considered in sect.~\ref{eft}. Furthermore, we introduce the neutrino portal model and its realization in both $s$- and $t$-channel annihilation scenarios in sect.~\ref{nuportal}. The $U(1)_{L_\mu-L_\tau}$ gauge model is discussed in sect.~\ref{u1} in the context of model parameters fixed by the measurements of the anomalous magnetic moment of the muon, which give the observed DM relic density. For all three models, we investigate the parameter space probed in the present and future neutrino experiments. We then summarize our results in sect.~\ref{summary}.

\section{Spin-$\frac{3}{2}$ framework}
\label{framework}
\begin{figure*}[t]
  \centering
  \subfloat[\label{fig1s}]{\includegraphics[width=0.45\textwidth]{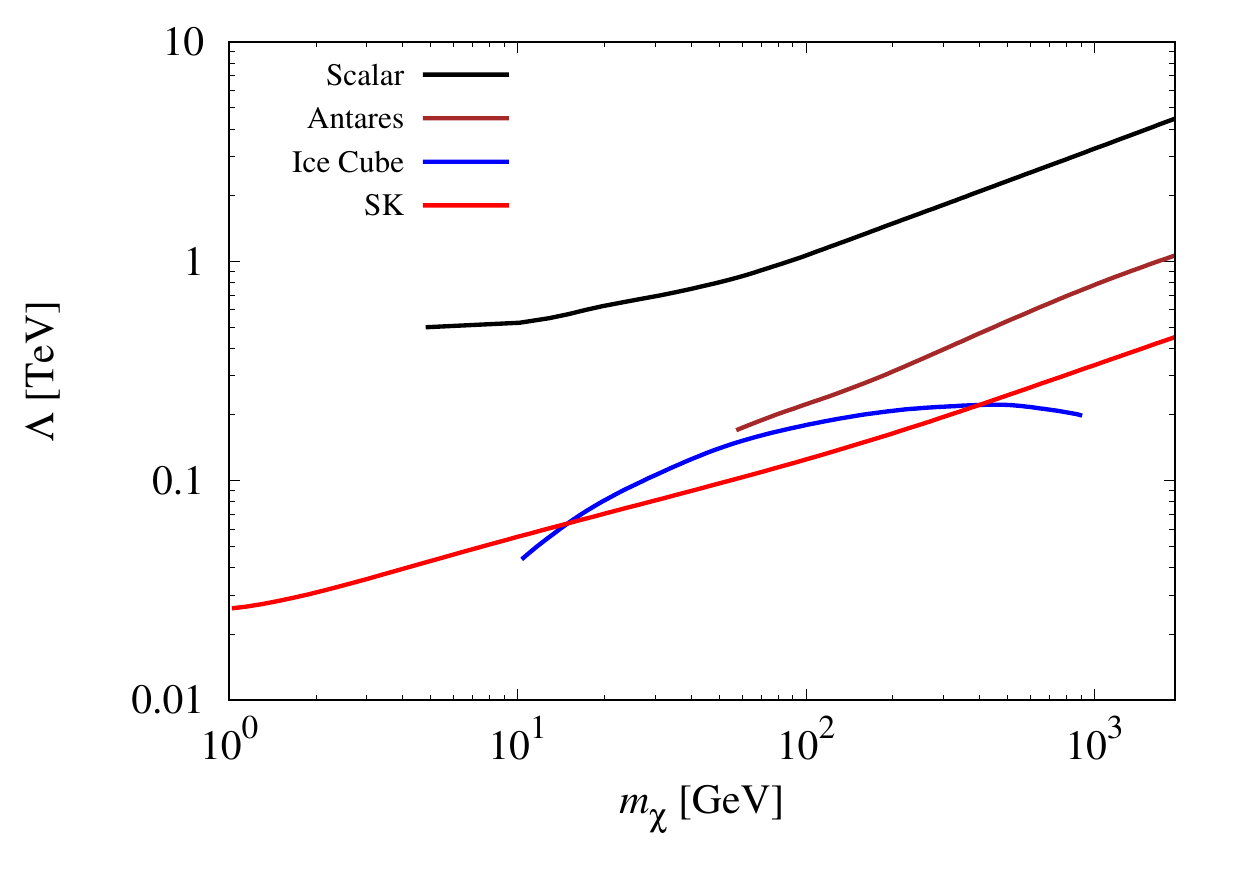}}
  \subfloat[\label{fig1v}]{\includegraphics[width=0.45\textwidth]{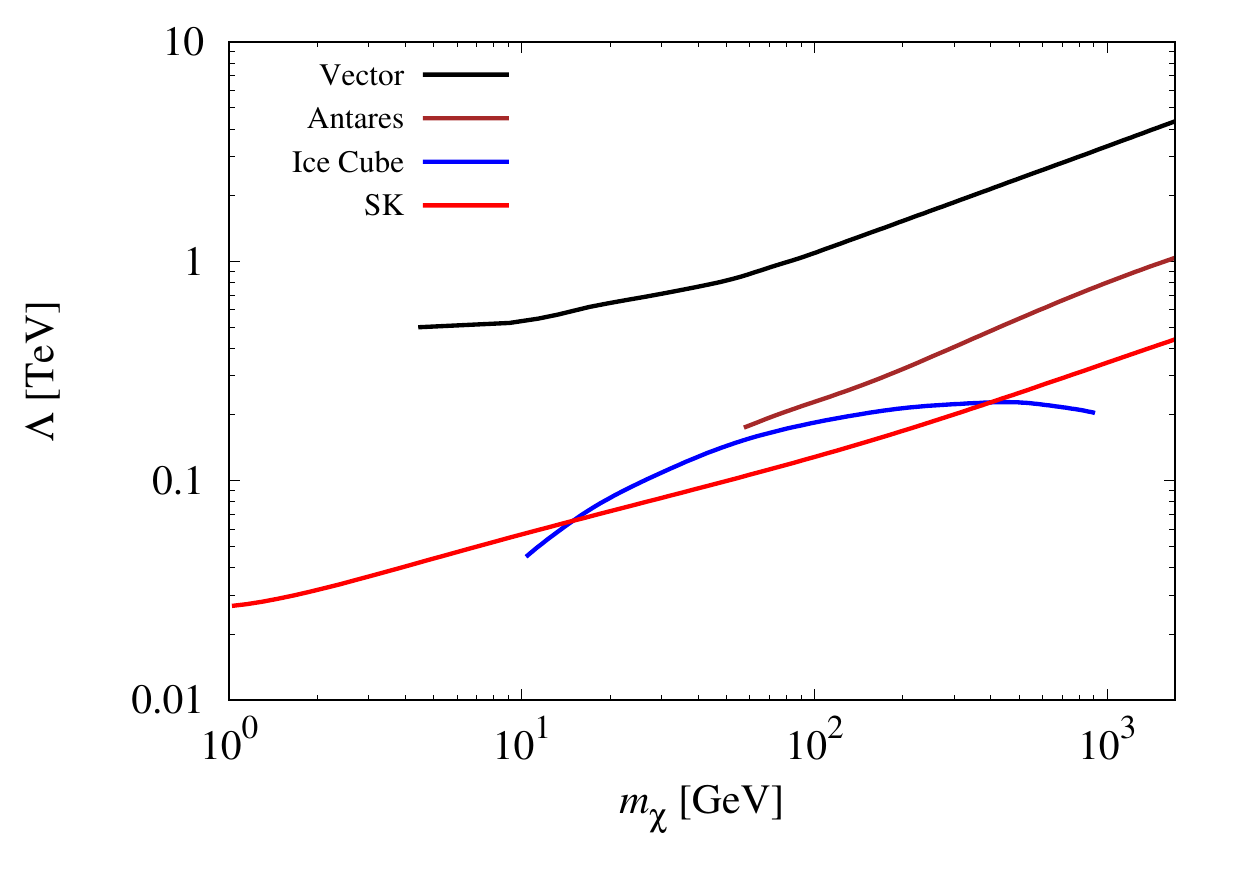}} \\
  \subfloat[\label{fig1t1}]{\includegraphics[width=0.45\textwidth]{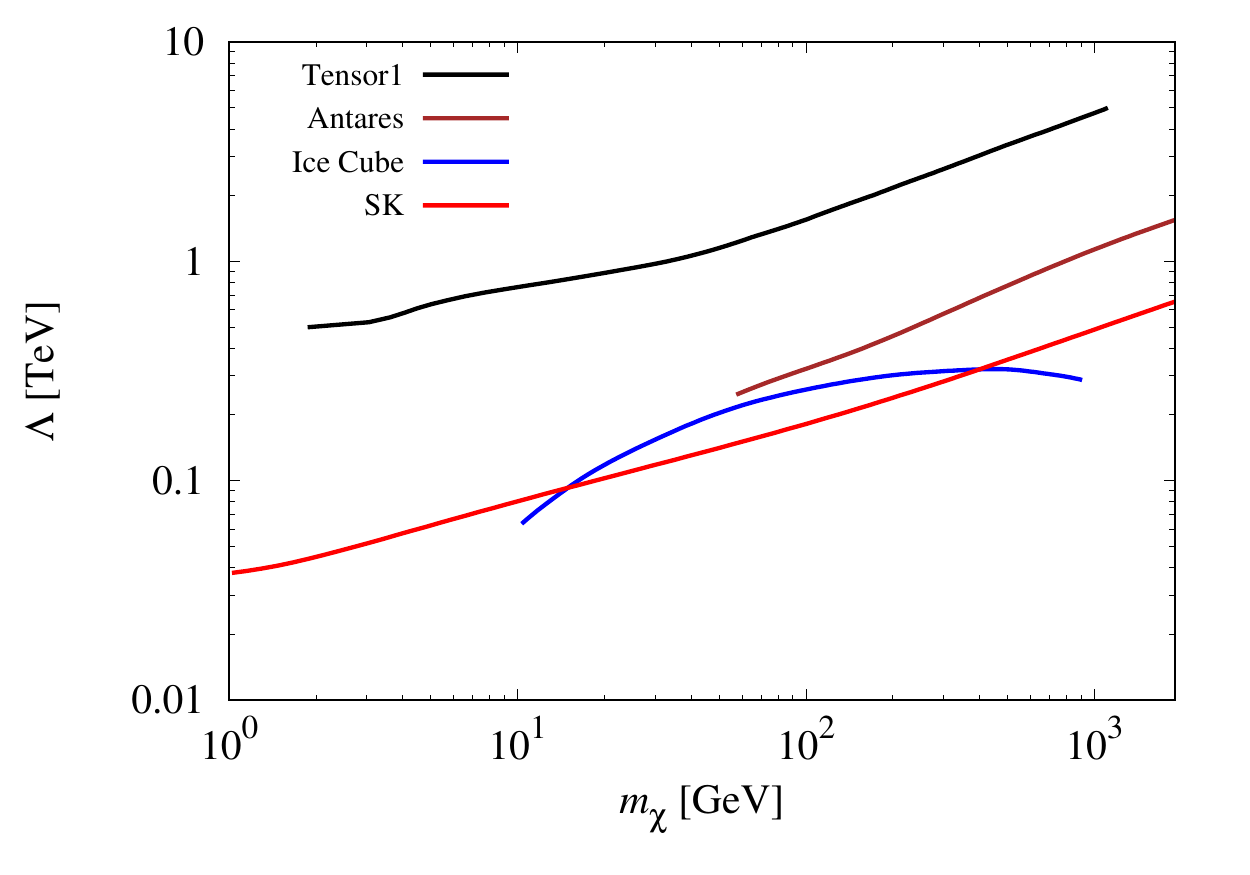}}
  \subfloat[\label{fig1t2}]{\includegraphics[width=0.45\textwidth]{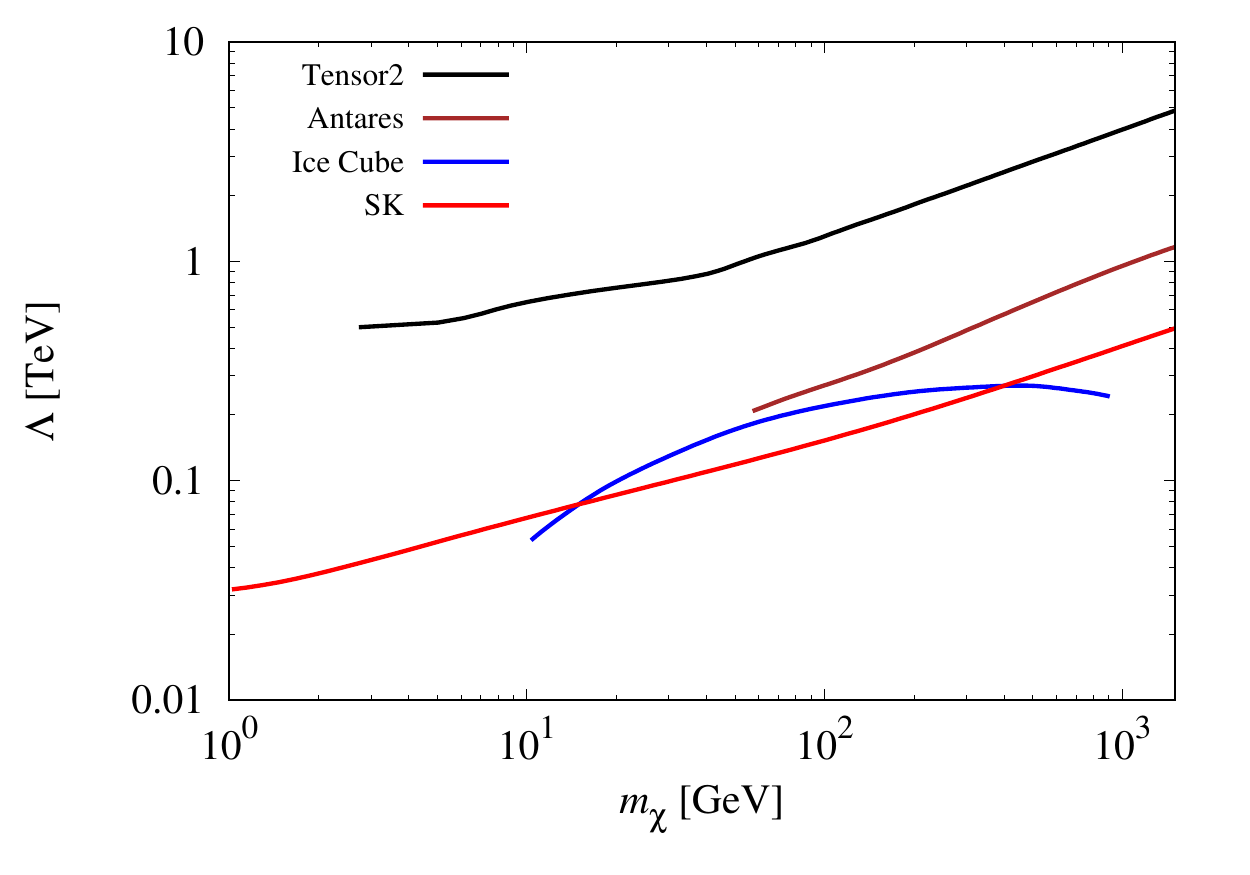}}
  \caption{\small Constraints on the DM mass $m_\chi$ and the cut-off scale $\Lambda$ for the (a) scalar, (b) vector, (c) Tensor1 and (d) Tensor2 interactions. The solid black lines in all four panels are plotted under the assumption that the spin-$\frac{3}{2}$ DM, $\chi_\mu$, saturates the observed relic density $\Omega_{\rm DM} h^2 \simeq 0.12$. The brown, blue and red solid lines show the constraints from Antares, Ice Cube and SK experiments.   }
\label{fig1}
\end{figure*} 

A massive spin-$\frac{3}{2}$ field $\chi_\mu$ with mass $m_\chi$ is described by the Lagrangian~\cite{Christensen:2013aua,Stirling:2011ya}:
\begin{equation}
    {\mathcal{L}}_{\frac{3}{2}} = -\frac{1}{2} {\bar{\chi}}_\mu \left(i \gamma^{\mu\rho\nu} \partial_\rho + m_\chi \gamma^{\mu\nu} \right)\chi_\nu, \label{l32}
\end{equation}
where $\gamma^{\mu\rho\nu} = -i \epsilon^{\mu\rho\nu\sigma}\gamma_5\gamma_\sigma$ and $\gamma^{\mu\nu} = \frac{1}{2}\left[\gamma^\mu,\gamma^\nu \right]$. The equation of motion for $\chi_\mu$ is given by
\begin{equation}
    i\gamma^{\mu\nu\rho}\partial_\nu\chi_\rho + m_\chi \gamma^{\mu\nu}\chi_\nu = 0, \qquad \gamma^{\mu\nu}\partial_\mu\chi_\nu = 0,
\end{equation}
along with the on mass-shell condition $\gamma^\mu\chi_\mu = \partial^\mu \chi_\mu = 0$. The equation of motion reduces to
\begin{equation}
    \left(-i \slashed{\partial} + m_\chi\right) \chi_\mu = 0.
\end{equation}
The polarization sums for spin-$\frac{3}{2}$ particles are
\begin{equation}
    S^{+}_{\mu\nu} = \sum_{i=-3/2}^{3/2} u^i_\mu (p) {\bar{u}}^i_\nu (p); \quad S^{-}_{\mu\nu} = \sum_{i=-3/2}^{3/2} v^i_\mu (p) {\bar{v}}^i_\nu (p)
\end{equation}
and are given by
\begin{align}
    S^\pm_{\mu\nu}(p) = -\left( \slashed{p} \pm m_\chi\right) \Big[g_{\mu\nu} - \frac{1}{3}\gamma_\mu\gamma_\nu - \frac{2}{3 m^2_\chi} \slashed{p}_\mu \slashed{p}_\nu \notag\\
    \mp \frac{1}{3 m_\chi}\left(\gamma_\mu p_\nu - \gamma_\nu p_\mu \right) \Big].
\end{align}
Following this short introduction of the spin-$\frac{3}{2}$ field $\chi_\mu$, in further sections we will consider $\chi_\mu$ as a DM candidate in all models with mass $m_\chi$. 

\section{Neutrino-specific (EFT) model}
\label{eft}
Even though spin-$\frac{3}{2}$ field theories are non-renormalizable, one can write generic interactions respecting the SM gauge symmetry and consider an EFT below a certain cut-off scale $\Lambda\gg m_\chi$. Introducing light right-handed neutrino fields ($\nu_j$), the most general dimension-6 effective Lagrangian relevant for spin-$\frac{3}{2}$ DM ($\chi_\mu$) annihilation into neutrinos can be written as:
\begin{equation}
    {\mathcal{L}}_{\rm int} = \sum_{j=e,\mu,\tau}\,\,\,\sum_{i=S,V,T_1,T_2} \frac{c_i}{\Lambda^2} \Theta_i^{\nu_j},
\end{equation}
where,
\begin{align}
    &\Theta_S^{\nu_j} = {\bar{\chi}_\mu}\left(a_S + i  b_S\gamma_5\right)\chi^\mu\, {\bar{\nu}_j}\left(a_S^\prime + i  b_S^\prime\gamma_5\right)\nu_j, \\
    &\Theta_V^{\nu_j} = {\bar{\chi}_\mu}\left(a_V +  i b_V\gamma_5\right)\gamma^\alpha\chi^\mu\, {\bar{\nu}_j}\left(a_V^\prime + i  b_V^\prime\gamma_5\right)\gamma_\alpha\nu_j,\\
    &\Theta_{T_1}^{\nu_j} = {\bar{\chi}_\mu}\left(a_{T_1} + i b_{T_1}\gamma_5\right)\sigma_{\alpha\beta}\chi^\mu\, {\bar{\nu}_j}\left(a_{T_1}^\prime + i b_{T_1}^\prime\gamma_5\right)\sigma^{\alpha\beta}\nu_j,\\
    &\Theta_{T_2}^{\nu_j} = \left\{{\bar{\chi}_\mu}\left(a_{T_2} + i b_{T_2}\gamma_5\right)\chi_\nu - {\bar{\chi}_\nu}\left(a_{T_2}^\prime + i b_{T_2}^\prime\gamma_5\right)\chi_\mu \right\} \notag\\ &\qquad\qquad\cdot{\bar{\nu}_j}\left(a_{T_2}^\prime + i b_{T_2}^\prime\gamma_5\right)\sigma^{\mu\nu}\nu_j,
\end{align}
and the $c_i$'s are the overall corresponding coupling strengths for scalar ($S$), vector ($V$), Tensor1 ($T_1$) and Tensor2 ($T_2$) terms. 

In the non-relativistic limit, the DM annihilation cross-section $\sigma|\rm v|$ for the above four fermion effective interactions are given as:
\begin{align}
    (\sigma|{\rm v}|)^S \simeq &\,\frac{c_S^2 m_\chi^2}{4\pi\Lambda^4} \left\{ \left(a_S^\prime \right)^2 + \left(b_S^\prime \right)^2 \right\} \notag \\ 
    &\times \left[a_S^2 + \frac{{\rm v}^2}{72}\left(51 a_S^2+10b_S^2 \right) \right],
\end{align}
\begin{align}
    (\sigma|{\rm v}|)^V \simeq &\,\frac{5c_V^2 m_\chi^2}{18\pi\Lambda^4} \left\{ \left(a_V^\prime \right)^2 + \left(b_V^\prime \right)^2   \right\} \notag \\ 
    &\times \left[a_V^2 + \frac{{\rm v}^2}{24}\left(4 a_S^2+15 b_S^2 \right) \right],
\end{align}
\begin{align}
    (\sigma|{\rm v}|)^{T_1} \simeq &\, \frac{c_{T_1}^2 m_\chi^2}{9\pi\Lambda^4}\left( a_{T_1}^2 + b_{T_1}^2 \right) \notag \\ 
    &\times  \left\{ \left(a_{T_1}^\prime \right)^2 + \left(b_{T_1}^\prime \right)^2   \right\}  \left[5 + \frac{83 {\rm v}^2}{24} \right],
\end{align}
and
\begin{align}
    (\sigma|{\rm v}|)^{T_2} \simeq &\,\frac{c_{T_2}^2 m_\chi^2}{9\pi\Lambda^4}\left\{ \left(a_{T_2}^\prime \right)^2 + \left(b_{T_2}^\prime \right)^2   \right\} \notag \\ 
    &\times \left[5 a_{T_2}^2 + \frac{{\rm v}^2}{24} \left(93 a_{T_2}^2 + 22 b_{T_2}^2 \right) \right].
\end{align}
Further, we calculate the corresponding relic density by numerically solving the Boltzmann equation:
\begin{align}
    \frac{d\eta_\chi}{dt} + 3 H \eta_\chi = - \langle\sigma|{\rm v}\rangle \left(\eta_\chi^2 - (\eta_\chi^{\rm eq.})^2 \right),
\end{align}
where $\langle\sigma|{\rm v}\rangle$ is the thermally averaged $\chi_\alpha$-annihilation cross section and $\eta_\chi$ is the number density of $\chi_\alpha$'s~\cite{Kolb:1990vq,Beltran:2008xg}. In Fig.~\ref{fig1}, we show the contour graphs in the $m_{\chi}$-$\Lambda$ plane that conform with the observed relic density $\Omega_{\text{DM}} h^{2} \simeq 0.12$.
\begin{table}[t]
\begin{center}
      \begin{tabular}{lc}
      \hline
\bf Experiment & \bf Energy-range \\ \hline
Hyper-Kamiokande (HK)~\cite{Hyper-Kamiokande:2018ofw} & \\
sensitivity one order of & \\
magnitude greater than in & \\
Super-Kamiokande (SK)~\cite{Super-Kamiokande:2020sgt, Frankiewicz:2015zma}  & $1-10^4$~{\rm GeV} \\ \hline
Ice-Cube~\cite{IceCube:2016oqp}, IC-Upgrade~\cite{Baur:2019jwm} & $20-10^4$~{\rm GeV} \\ \hline
Antares~\cite{ANTARES:2015vis} & $50-10^5$~{\rm GeV} \\ \hline
DUNE~\cite{DUNE:2015lol} & $25-100$~{\rm MeV} \\ \hline
\end{tabular}
      \caption{Existing and upcoming neutrino experiments measuring neutrino flux in different energy ranges.}
      \label{tab1}
      \end{center}
\end{table}
All couplings are set equal to one and $\sigma|\rm v|$ is summed over all  neutrino flavors. The existing and upcoming neutrino experiments~\cite{Hyper-Kamiokande:2018ofw,Super-Kamiokande:2020sgt,Frankiewicz:2015zma,IceCube:2016oqp,Baur:2019jwm,ANTARES:2015vis,DUNE:2015lol} in different energy ranges are given in Table~\ref{tab1}. Their contributions to the neutrino flux are translated into the constraints on ($m_\chi, \Lambda$) parameters. The solid black line in Fig.~\ref{fig1} depicts the contours for the observed relic density $\Omega_{\rm DM} h^2 \simeq 0.12$ for the scalar (Fig.~\ref{fig1s}), vector (Fig.~\ref{fig1v}), Tensor1 (Fig.~\ref{fig1t1}) and Tensor2 (Fig.~\ref{fig1t2}) couplings in the EFT framework. The brown, blue and red solid lines depict the constraints from the observed or prospective sensitivities from the existing and upcoming experiments. The contours in Fig.~\ref{fig1} show that in EFT framework the new physics scale $\Lambda \in [\sim 0.5 - 5]$~TeV provides the correct observed relic density for $m_\chi$ lying between a few GeV to the TeV range. However, the constraints on $\Lambda$ from the experiments ($\Lambda \in \sim[10 - 800]$~GeV) are far below the required valid range for the EFT framework.

\section{Neutrino-portal model}
\label{nuportal}
Following a simple model considered in Refs.~\cite{Batell:2017cmf,Blennow:2019fhy}, in which DM couples to the SM neutrinos through a mixing generated by a sterile pseudo-Dirac neutrino, $N$, we can consider the interaction Lagrangian as
\begin{equation}
    -{\mathcal{L}} \supset \lambda_\ell {\bar{L}}_\ell \Tilde{H} N_R + {\mathcal{L}}_{SM} + {\bar{N}} \left(i {\slashed{\partial}} - m_N \right) N,
\end{equation}
where $N_R$ is the right-handed projection of the sterile neutrino $N$.
After electroweak symmetry breaking, the SM neutrinos $\nu_\alpha$ $(\alpha = e, \mu, \tau)$ mix with $N$ through $(\lambda_\ell {\rm v}_0/\sqrt{2})\bar{\nu}_{\alpha L} N_R$, where ${\rm v}_0 = 246$~GeV is the Higgs vacuum expectation value. Diagonalization of the mass mixing leads to a heavy sterile Dirac neutrino state $\nu_4$ with mass 
\begin{equation}
m_4 = \left(m_N^2 + \sum_{\ell=e,\mu,\tau} \frac{\lambda_\ell^2 {\rm v}_0^2}{2}\right)^{1/2}.
\end{equation}
Note that the lepton number symmetry forbids active SM neutrinos from acquiring mass. In order to account for the neutrino mass, small lepton number breaking through a term $\mu_N {\bar{N}}_L N_L^C$, via an inverse see-saw mechanism~\cite{Mohapatra:1986bd,Bernabeu:1987gr}, is required. The neutrino mixing matrix $U$ relates left-handed flavor neutrino fields $N_L$ with the neutrino mass-eigenstates $\nu_{\alpha L}$ as
\begin{equation}
    \begin{pmatrix} \nu_{\alpha L} \\ N_L \end{pmatrix} = U \begin{pmatrix} \nu_{i L} \\ \nu_{4 L} \end{pmatrix},\quad {\rm with} \quad U = \begin{pmatrix}
        U_{\alpha i} & U_{\alpha4} \\
        U_{si} & U_{s4}
    \end{pmatrix}.
\end{equation}
Here $U_{\alpha i}$ corresponds to $3\times 3$ PMNS matrix and the mixing elements can be written as 
\begin{align}
    &U_{\alpha4} = \frac{\theta_\alpha}{\sqrt{1+\sum_\alpha |\theta_\alpha|^2}};\quad
    U_{s4} = \frac{1}{\sqrt{1+\sum_\alpha |\theta_\alpha|^2}}; \notag\\
    &\sum_{i=1}^3 \left|U_{si} \right|^2 = \sum_{\alpha = e, \mu, \tau} \left|U_{\alpha4} \right|^2. \label{mixang}
\end{align}
The detailed description of this formalism can be found in Refs.~\cite{Batell:2017cmf,Blennow:2019fhy}. For definiteness, as in Ref.~\cite{Fernandez-Martinez:2016lgt}, the mixing angles $\theta_\alpha = \lambda_\alpha {\rm v}_0/(\sqrt{2} m_N)$ are fixed to $|\theta_e| = 0.031$, $|\theta_\mu| = 0.011$ and $|\theta_\tau| = 0.044$. 

We implement this neutrino-portal model in the $s$- and $t$-channel DM ($\chi_\alpha$) annihilation scenarios, where $\chi_\alpha$ is a SM singlet spin-$\frac{3}{2}$ fermion. In the $s$- ($t$-) channel, $\chi_\alpha$ interacts with the SM neutrinos through the mixing of sterile neutrinos via a vector (scalar or vector) mediator.
 
\subsection{$s$-channel mediator model}
\label{schannel}
The Lagrangian ${\mathcal{L}}_{\rm int}$ relevant for the interactions between $\chi_\alpha$ and the vector-boson $Z^\prime_\mu$, which mediates the interaction between the dark and the visible sector, can be written as
\begin{equation}
    {\mathcal{L}}_{\rm int} \supset g_{Z^\prime} {\bar{\chi}}_\mu \gamma^\nu \chi^\mu Z^\prime_\nu + g_N {\bar{N}}_L \gamma^\nu N_L Z^\prime_\nu. \label{lzp}
\end{equation}
This equation can be described by a $U(1)^\prime$ gauge symmetry being spontaneously broken by the vacuum expectation value of a $U(1)^\prime$ charged scalar singlet~\cite{Blennow:2019fhy,Batell:2017cmf}. The new vector mediator $Z^\prime_\nu$ connects DM with the active neutrinos through the mixing of the sterile Dirac neutrino with the active SM neutrinos. It would then generate masses for $Z^\prime_\nu$, $N$ and the DM $\chi_\mu$. A $\mathbb{Z}_2$ symmetry would prevent mixing between the neutrinos and the DM. Here the $U(1)^\prime$ charges of DM and the sterile neutrino are assumed to be equal, which leads to $g_{Z^\prime} = g_N$. The DM can also annihilate into charged leptons through the loop induced coupling of $Z^\prime_\nu$, through the kinetic mixing of $Z^\prime_\nu$ with the $Z$-boson, as discussed in Ref.~\cite{Holdom:1985ag}. It has been shown in Ref.~\cite{Blennow:2019fhy} that the DM annihilation cross-section to charged leptons is highly suppressed, being several orders of magnitude smaller compared to the DM annihilation into SM neutrinos. However for $m_\chi < m_4$ and $m_\chi < m_{Z^\prime}$, the DM annihilation channel into three SM neutrinos is dominant. 
\begin{figure*}[t]
  \centering
  \subfloat[\label{fig2a}]{\includegraphics[width=0.45\textwidth]{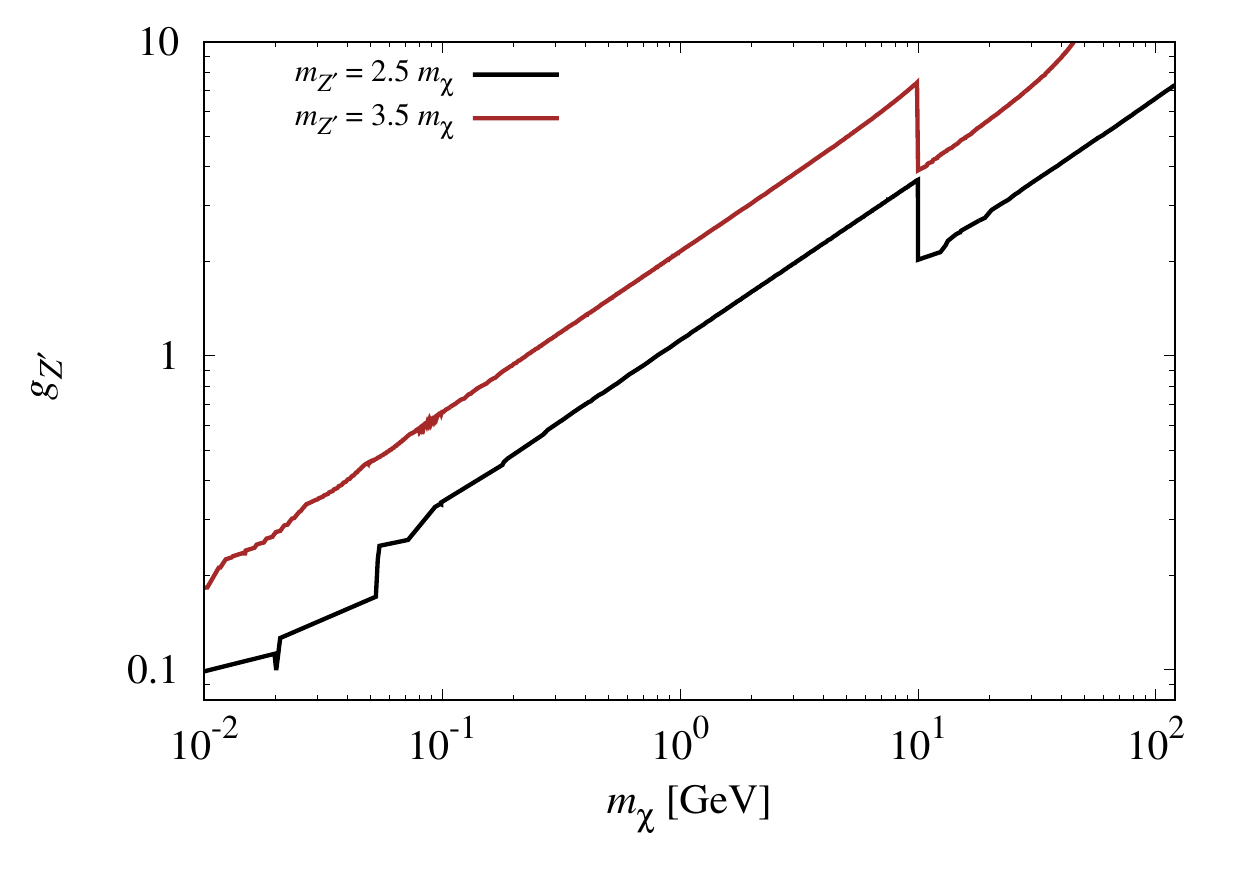}}
  \subfloat[\label{fig2b}]{\includegraphics[width=0.45\textwidth]{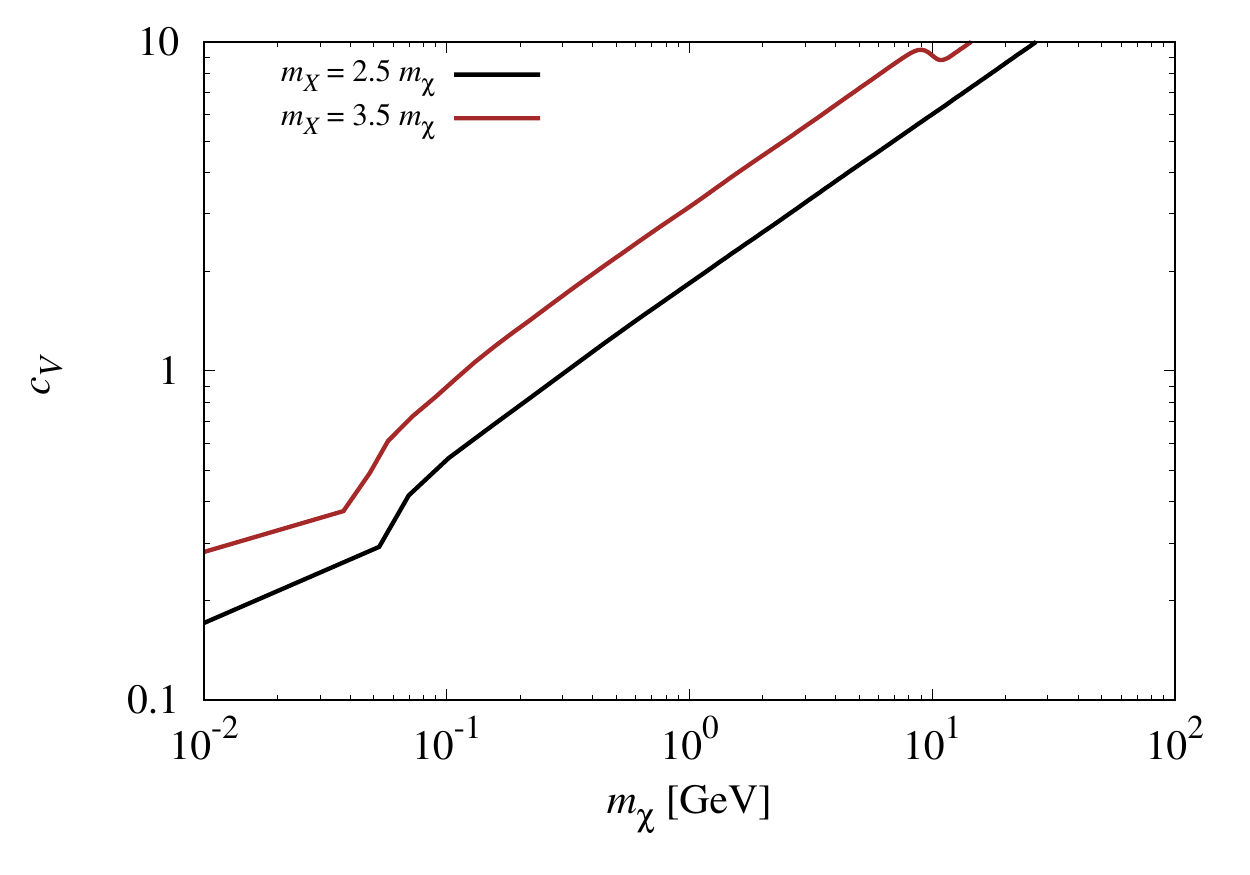}} \\
  \subfloat[\label{fig2c}]{\includegraphics[width=0.45\textwidth]{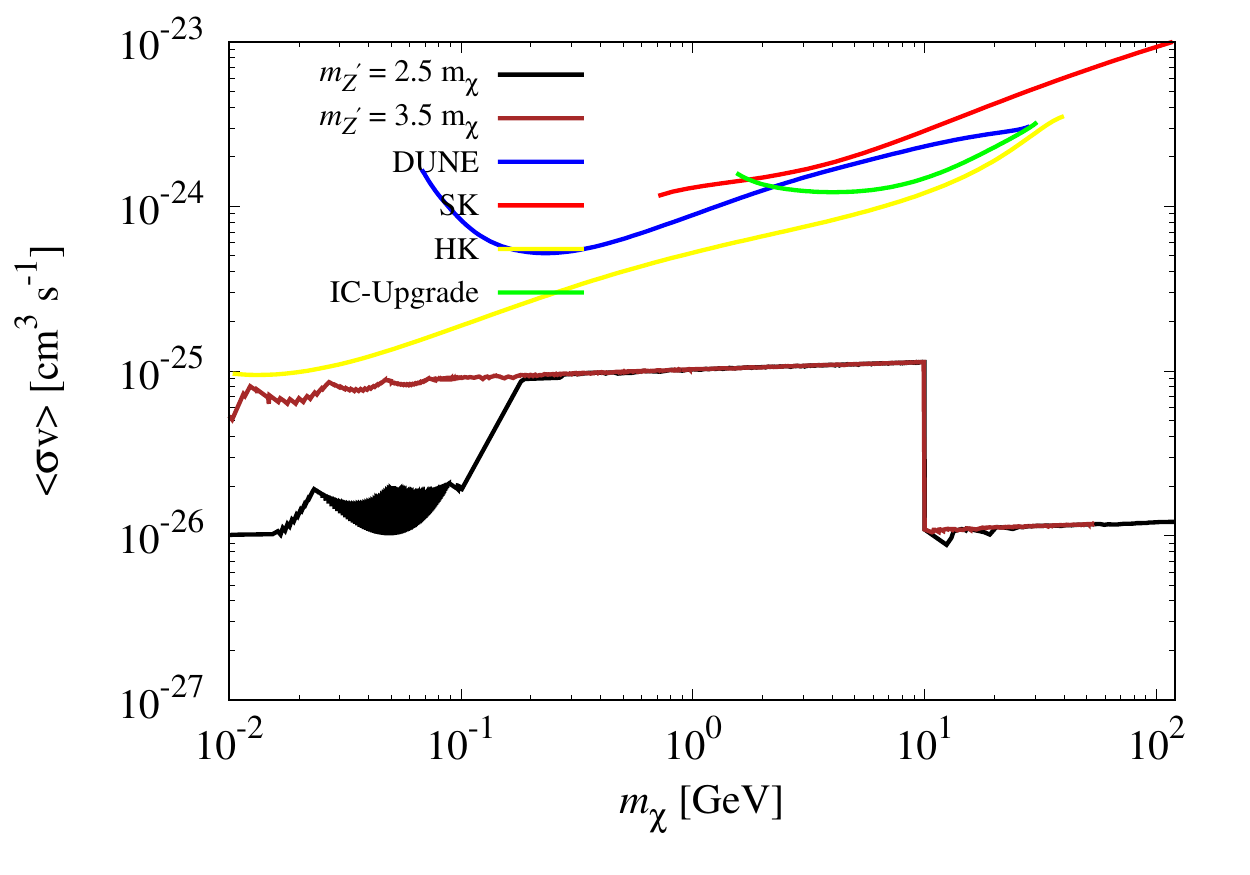}}
  \subfloat[\label{fig2d}]{\includegraphics[width=0.45\textwidth]{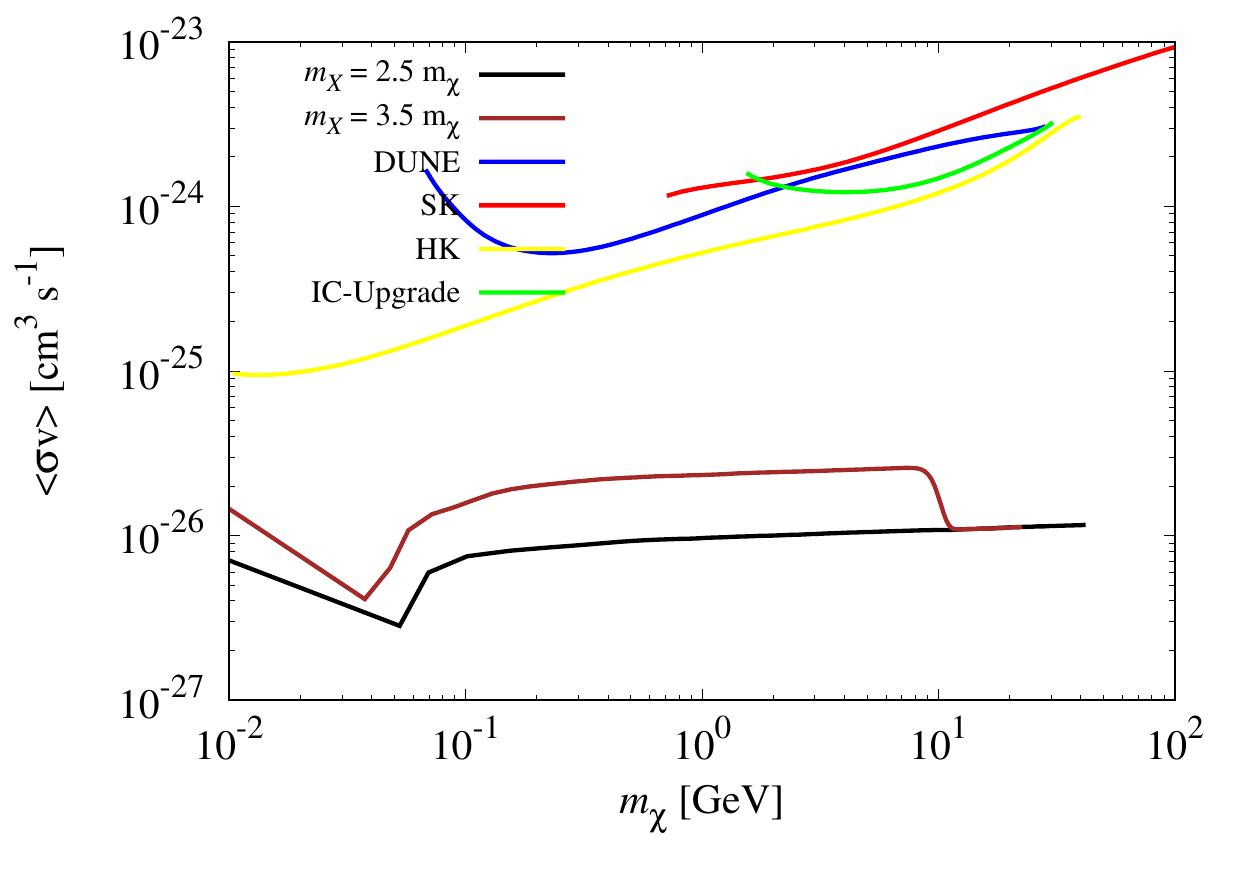}}
  \caption{\small Relic density and thermally averaged DM annihilation cross-section $\langle\sigma {\rm v}\rangle$ plots in the neutrino-portal model. The top panel shows the relic density plots in the $(m_\chi - g_{Z^\prime}/c_V)$ plane assuming that the relic density is saturated by the DM annihilation. The left and the right panels are for the $s$- and $t$-channel 4D vector mediation interactions. In the lower panel we show the constraints on $\langle\sigma {\rm v}\rangle$ from the observed or prospective sensitivities from DUNE (solid blue), SK (red), HK (yellow) and IC-upgrade (green) experiments, and compare them with the thermally averaged cross sections of the vector couplings ($g_{Z^\prime}, c_V$) which are fit to give the observed relic density (shown in the upper panel).}
  \label{fig2}
\end{figure*} 

In this model the thermally averaged DM annihilation cross-section into SM neutrinos is given by
\begin{align}
    \langle \sigma|\rm v| \rangle &= \frac{5 g_{Z^\prime}^4}{144 \pi} \left(\sum_{\alpha = e, \mu, \tau} \left|\theta_\alpha\right|^2 \right)^2 \frac{m_\chi^2}{\left( 4 m_\chi^2 - m_{Z^\prime}^2\right)^2 + \Gamma^2 m_{Z^\prime}^2},
\end{align}
where the decay width $\Gamma$ of $Z^\prime_\mu$ is given by
\begin{align}
    \Gamma& =\frac{g_{Z^\prime}^{2}}{12\pi}m_{Z^{\prime}} \left(\sum_{\alpha = e, \mu, \tau} \left|\theta_\alpha\right|^2\right)^2 \notag\\ 
    &+ \frac{g_{Z^\prime}^{2}}{12\pi}m_{Z^{\prime}}\Bigg[\frac{1}{36}\sqrt{1-  4\left(\frac{m_{\chi}}{m_{Z^{\prime}}}\right)^{2}} \cdot\left(\frac{m_{\chi}}{m_{Z^{\prime}}}\right)^{2}\notag\\
    &\Bigg(-4 + 24\left(\frac{m_{Z^{\prime}}}{m_{\chi}}\right)^{2} - 10\left(\frac{m_{Z^{\prime}}}{m_{\chi}}\right)^{4} + 2\left(\frac{m_{Z^{\prime}}}{m_{\chi}}\right)^{6}
    \Bigg) \Bigg]. \label{decay1}
\end{align}
In eq.~(\ref{decay1}), the first term indicates the decay width of $Z^\prime \to \nu_\alpha\bar{\nu}_\alpha$ and the second term is for $Z^\prime \to \chi_\mu \bar{\chi}_\mu$. For a valid perturbative description of $Z^\prime_\nu$ as the mediator, the total decay width $\Gamma < m_{Z^\prime}$. The mediator $Z^\prime_\nu$ decays into SM neutrinos through the mixing of sterile and SM neutrinos with the mixing angles given in eq.~(\ref{mixang}). In addition, if the DM mass is less than $m_{Z^\prime}/{2}$, the mediator can decay into DM pairs. In this case, because of the spin-$\frac{3}{2}$ nature of DM, there also exists a minimum $m_\chi$ below which the perturbative analysis is not valid. The two limits are given as
\begin{align}
    \frac{g_{Z^\prime}^2}{12 \pi}\left(\sum_{\alpha = e, \mu, \tau} \left|\theta_\alpha\right|^2 \right)^2 < 1,\quad 
    \frac{g_{Z^\prime}^2}{216 \pi}\left(\frac{m_{Z^\prime}}{m_\chi} \right)^4 < 1. \label{limits}
\end{align}
The first inequality arises from the mediator $Z^\prime_\nu$ decaying into SM neutrinos, and the second inequality arises from its decay into $\chi_\mu$.

\subsection{$t$-channel mediator model}
\label{tchannel}
In the $t$-channel mediator model, the mediator can be a scalar or a vector.
\begin{itemize}
    \item[(a)] Scalar mediator $S$:
\end{itemize}
For the scalar mediator case, we can write the SM gauge invariant interaction between spin-$\frac{3}{2}$ $\chi_\mu$, spin-$\frac{1}{2}$  sterile Dirac neutrino $N_L$ and the scalar $S$ as~\cite{Pascalutsa:1994tp}:
\begin{equation}
    {\mathcal{L}}_{\rm int} \supset \frac{g_\chi}{\Lambda} {\bar{\chi}}_\mu \Theta^{\mu\nu} \left(\partial_\nu S\right) N_L + {\rm h.c.}, \label{22}
\end{equation}
where $\Theta^{\mu\nu}$ is taken to be $g^{\mu\nu}-\gamma^\mu\gamma^\nu$ and $g_\chi$ is the corresponding coupling strength. For $\chi_\mu$ on mass-shell
\begin{equation}
    {\mathcal{L}}_{\rm int} = \frac{g_\chi}{\Lambda} {\bar{\chi}}_\mu \left(\partial^\mu S\right) N_L + {\rm h.c.}. \label{23}
\end{equation}
In this case there will be no dimension-4 (4D) interaction term and we assume that the scalar field $S$ does not mix with the SM Higgs-boson. 
Therefore, the thermally averaged DM annihilation cross-section $\langle \sigma |\rm v|\rangle$ is given by
\begin{equation}
    \langle\sigma|{\rm v}|\rangle = \frac{g_\chi^4}{9216 \Lambda^4} \left(\sum_{\alpha = e, \mu, \tau} \left|\theta_\alpha \right|^2\right)^2 \frac{m_\chi^6}{\left(m_\chi^2 + m_S^2 \right)^2}, \label{24} 
\end{equation}
where $m_S$ is the mass of a scalar mediator field $S$.
\begin{itemize}
    \item [(b)] Vector-mediator $X_\mu$:
\end{itemize}
In this case we can write the 4D as well as dimension-5 (5D) interaction terms as
\begin{equation}
    {\mathcal{L}}_{\rm int}^{\rm 4D} = i c_V {\bar{\chi}}_\mu P_R N X^\mu + {\rm h.c.}, \label{25}
\end{equation}
and 
\begin{equation}
    {\mathcal{L}}_{\rm int}^{\rm 5D} = i \frac{g_V}{\Lambda} {\bar{\chi}}_\mu \gamma_\nu P_R N X^{\mu\nu} + {\rm h.c.}, \label{26}
\end{equation}
respectively, where $X^{\mu\nu} = \partial^\mu X^\nu - \partial^\nu X^\mu$ and $c_V$  $(g_V)$ is the coupling strength for the 4D (5D) interaction. The corresponding thermally averaged DM annihilation cross-sections are given as
\begin{align}
    \langle\sigma|{\rm v}|\rangle^{\rm 4D} = \frac{c_V^4}{1152 \pi} &\left(\sum_{\alpha = e, \mu, \tau} \left|\theta_\alpha \right|^2\right)^2 \notag\\
    &\times \left(\frac{2 m_{\chi}^2}{m_X^4}+\frac{3 m_{\chi}^2}{(m_{\chi}^2+m_X^2)^2} \right), \label{27}
\end{align}
and
\begin{equation}
    \langle\sigma|{\rm v}|\rangle^{\rm 5D} = \frac{5 g_V^4}{18432 \pi \Lambda^4} \left(\sum_{\alpha = e, \mu, \tau} \left|\theta_\alpha \right|^2\right)^2 \frac{m_\chi^6}{\left(m_\chi^2 + m_X^2\right)^2}, \label{28}
\end{equation}
where $m_X$ is the mass of the vector mediator $X_\mu$.  In the $t$-channel mediator model, the DM field $\chi_\mu$ and the mediators $S$ and $X_\mu$, forming the dark sector, are SM singlets. The Lagrangian respects a global $U(1)_{\mathbb{Z}}$ symmetry under which the mediators, sterile neutrino and the DM have the same charge. The Lagrangian also respects a global $U(1)_d$ dark symmetry under which the DM and mediators have equal charge. The DM is stable for $m_\chi$ smaller than the mediator mass. Similar to the $s$-channel case, the contribution to the DM annihilation into charged leptons arises at the loop level and has been shown to be many orders of magnitude smaller than the annihilation into SM neutrinos~\cite{Blennow:2019fhy}.

In this model, for the 5D-interactions (eqs.~(\ref{22}) and~(\ref{26})) of scalar and vector mediators, the thermally averaged DM annihilation cross-sections (eqs.~(\ref{24}) and~(\ref{28})) are suppressed by the cut-off scale $\Lambda^4$. For $\Lambda \sim 1$~TeV, the thermally averaged DM annihilation cross-section is several orders of magnitude smaller than the cross-section required for obtaining the observed relic density through freeze-out. We will thus consider only the 4D-interaction in both the $s$- and $t$-channel to obtain the relevant parameter space for the observed relic density.  

In Fig.~\ref{fig2}, the upper panels show the results of the observed relic density in the ($m_\chi$-$g_{Z^\prime}$) and ($m_X$-$c_V$) plane for the $s$- and $t$-channel 4D-interactions with neutrinos (through the vector mediators $m_{Z^\prime}$ and $m_X$, respectively). For numerical computations we have kept the ratio of mediator mass to $m_\chi$ fixed at 2.5 and 3.5, shown by solid black and brown lines respectively. The relic density is computed by summing the DM annihilation cross-sections over all neutrino flavors, with the mixing angles given in Ref.~\cite{Fernandez-Martinez:2016lgt} and $m_\chi$ in the 10~MeV $\sim$ 100~GeV range. The lower bound on $m_\chi$ is set by the CMB and Big Bang Nucleosynthesis (BBN) constraints~\cite{Olivares-DelCampo:2017feq}. In the lower panels we have shown the constraints on $\langle\sigma {\rm v}\rangle$ from the observed or prospective sensitivities of existing and upcoming experiments. The solid blue, red, yellow and green lines show the constraints from DUNE~\cite{DUNE:2015lol}, SK~\cite{Super-Kamiokande:2020sgt,Frankiewicz:2015zma}, HK~\cite{Hyper-Kamiokande:2018ofw} and IC-upgrade~\cite{Baur:2019jwm}, respectively. In the freeze-out scenario the DUNE experiment will be able to exclude $25 \lesssim m_\chi \lesssim 100$~MeV. 

\section{$U(1)_{L_\mu-L_\tau}$ gauge model}
\label{u1}
We explore the $U(1)_{L_\mu-L_\tau}$ gauge model in the context of obtaining the observed DM relic density parameter space which explains the anomalous muon magnetic moment measurement, given as~\cite{Aoyama:2020ynm} 
\begin{equation*}
    \Delta a_\mu = a_\mu^{\rm experiment} - a_\mu^{\rm SM} = (251\pm 59) \times 10^{-11},
\end{equation*}
where $a_\mu$ $\equiv \frac{1}{2} (g-2)_\mu$. 

\begin{figure}[t]
\centering
\includegraphics[width=0.45\textwidth]{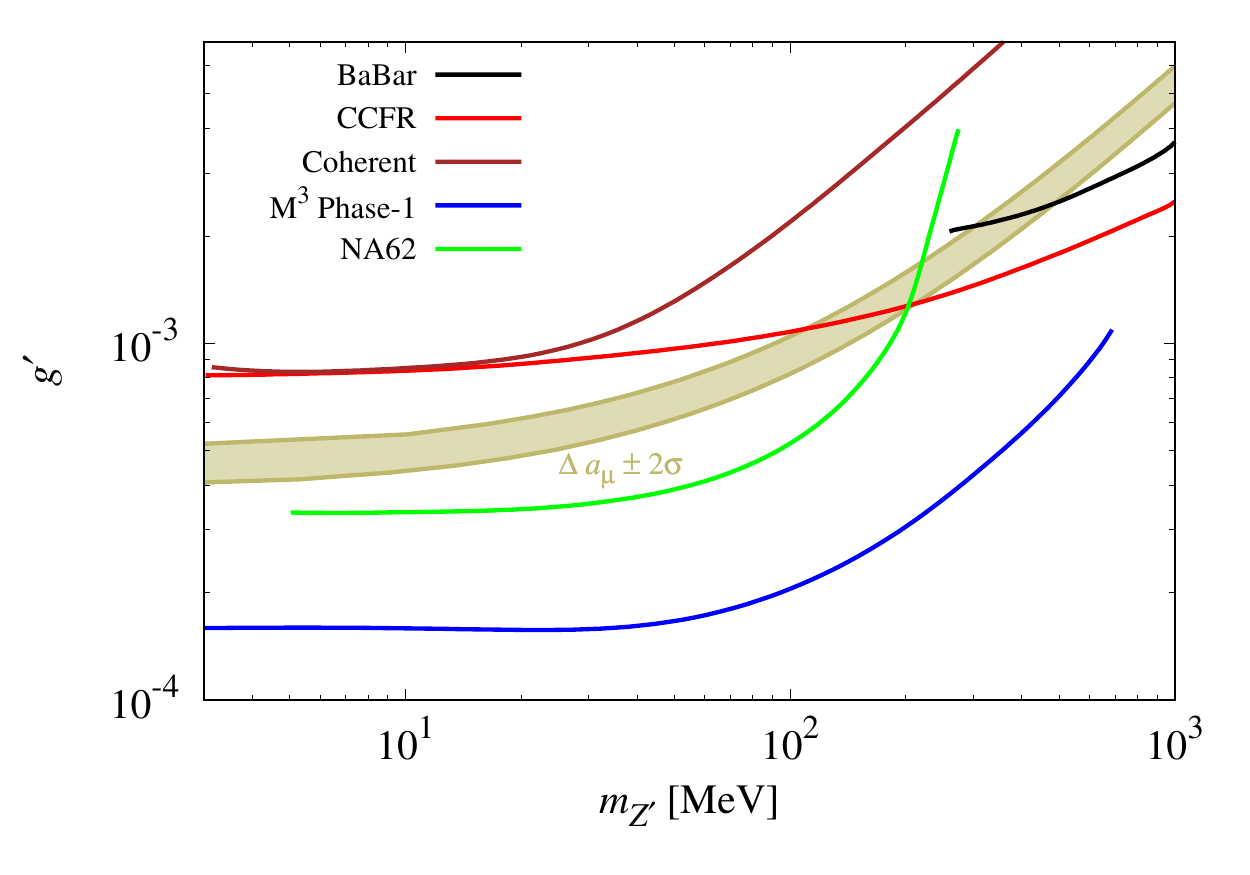} 
\caption{\small The colored band represents the regions of parameter space in which an $L_{\mu}-L_{\tau}$ gauge boson can resolve the $\Delta a_\mu$ anomaly (within $2\sigma$). Constraints from BaBar~\cite{BaBar:2016sci}, CCFR~\cite{CCFR:1991lpl,Altmannshofer:2014pba}, Neutrino Nucleus Coherent elastic scattering~\cite{Cadeddu:2020nbr}, $M^3$ phase-I~\cite{Kahn:2018cqs} and NA62~\cite{Krnjaic:2019rsv} experiments are shown by solid colored as indicated.}
\label{fig:amu}
\end{figure}

In this model there is a $U(1)_{L_\mu-L_\tau}$ gauge symmetry with a corresponding gauge boson $Z^\prime_\mu$ under which $\chi_\mu$ is a singlet and carries a $U(1)_{L_\mu-L_\tau}$ charge. The Lagrangian of the model is given as
\begin{equation}
    {\mathcal{L}}_{\rm int} \supset g^\prime\left( J^\mu_{L_\mu-L_\tau} +  J^\mu_{\rm DM} \right)Z_\mu^\prime,
\end{equation}
where $g^\prime$ is the universal coupling. The current
\begin{equation}
    J^\mu_{L_\mu-L_\tau} = {\bar{\mu}} \gamma^\mu \mu + {\bar{\nu_\mu}} \gamma^\mu \nu_\mu - {\bar{\tau}} \gamma^\mu \tau - {\bar{\nu_\tau}} \gamma^\mu \nu_\tau
\end{equation}
and the $U(1)_{L_\mu-L_\tau}$ charge are taken to be $+1$ for $L_\mu$ and $-1$ for $L_\tau$.  
If $\chi_\mu$ carries $q_\chi$ units of $L_\mu - L_\tau$ charge,
the corresponding $J_{\rm DM}^\mu$ is given as~\cite{Christensen:2013aua}
\begin{equation}
    J^\mu_{\rm DM} = i q_\chi \epsilon^{\mu\nu\rho\sigma} {\bar{\chi}}_\nu \gamma_5 \gamma_\sigma \chi_\rho. \label{jdm}
\end{equation}
Under the on mass-shell condition, eq.~(\ref{jdm}) reduces to
\begin{equation}
    J^\mu_{\rm DM} = q_\chi {\bar{\chi}}_\nu \gamma^\mu \chi^\nu.
\end{equation}
To start we first explore the relevant parameter space in the ($g^\prime$-$m_{Z^\prime}$) plane where the measured $\Delta a_\mu$ can be explained. The leading contribution to $\Delta a_\mu$ arises at the one loop level and is given by
\begin{equation}
    \Delta a_\mu = \frac{g^\prime}{4\pi^2} \int_0^1 dx \frac{x(1-x)^2}{(1-x)^2+\frac{m^2_{Z^\prime}}{m^2_\mu}x}.
\end{equation}
The parameter space in the coupling $g^\prime$ and gauge boson mass $m_{Z^\prime}$ has been widely explored and constrained from several experiments~\cite{Bauer:2018onh}. The BaBar experiments~\cite{BaBar:2016sci} have searched for the $Z^\prime_\mu$ boson coupling to muons through the muon-pair production in $e^+e^- \to \mu^+\mu^- Z^\prime$, $Z^\prime\to\mu^+\mu^-$. This experiment is relevant for $m_{Z^\prime} \in [0.212, 10]$~GeV range. Neutrino Trident Production CCFR~\cite{CCFR:1991lpl,Altmannshofer:2014pba} experiments obtain an additional contribution in the model, and put stringent limits typically requiring the coupling $g^\prime$ to lie between $10^{-3} - 10^{-4}$ for the $Z^\prime_\mu$-boson mass in 10~MeV to 1~GeV range. The coherent elastic neutrino-nucleus scattering experiment~\cite{Cadeddu:2020nbr} puts constraints through the $Z^\prime_\mu$-$\gamma$ mixing. Constraints from projected sensitivities from the $M^3$ phase-I~\cite{Kahn:2018cqs} missing muon momentum experiment will test much of the parameter space. In the NA62 experiment~\cite{Krnjaic:2019rsv}, the sensitivity in $Z^\prime_\mu$ is obtained through their production in $K \to \mu\nu Z^\prime_\mu$ decays and this experiment is relevant for $m_{Z^\prime}\in[10, 200]$~MeV. In Fig.~\ref{fig:amu} we show the allowed parameter space satisfying the $\Delta a_\mu$ in the $U(1)_{L_\mu-L_\tau}$ gauge model constrained by existing and future experiments. Clearly, much of the $\Delta a_\mu$ favored parameter space can be excluded by future NA62 and $M^3$ Phase-I experiments.

Using the values of $g^\prime$ and $m_{Z^\prime}$ which satsify $\Delta a_\mu$ at the $2\sigma$ level, we study the relic density constraints and other cosmological and astrophysical observations for $\chi_\mu$. In this model the DM annihilation cross-section to fermions is given as:
\begin{align}
    &\sigma_S\left( \chi_\mu \bar{\chi}_\mu \to f \bar{f}\right) \notag\\ &\,= \sum_f \frac{\kappa_f \left( g^\prime\right)^4}{432 \pi m_\chi^4 s} \sqrt{\frac{s-4m_f^2}{s-4m_\chi^2}} \frac{1}{\left(s- m_{Z^\prime}^2\right)^2 + \Gamma^2_{Z^\prime} m_{Z^\prime}^2} \notag\\
    &\quad\times \left[ \left(s+2m_f^2 \right) \left(36 m_\chi^6 - 2 m_\chi^4 s - 2 m_\chi^2 s^2 + s^3 \right) \right], 
\end{align}
where the sum runs over $f=\mu, \tau, \nu_\mu, \nu_\tau$ and $k_f = 1$ for $\mu, \tau$, and $\frac{1}{2}$ for $\nu_\mu, \nu_\tau$. 
The decay width $\Gamma_{Z^\prime}$ is given by
\begin{align}
    \Gamma_{Z^\prime}
    &=\, \sum_{f=\mu,\tau,\nu_{\mu},\nu_{\tau}}\frac{k_f {g^\prime}^{2}}{12\pi}\sqrt{1-\frac{4m_{f}^{2}}{m_{Z^{\prime}}^{2}}}m_{Z^{\prime}}\left(1+\frac{2m_{f}^{2}}{m_{Z^{\prime}}^{2}} \right)\notag\\
    &+\frac{k_{\chi} {g^\prime}^{2}}{108\pi}\sqrt{1-\frac{4m_{\chi}^{2}}{m_{Z^{\prime}}^{2}}} m_{Z^{\prime}}\left[\frac{m_{Z^{\prime}}^{4}}{m_{\chi}^{4}}+36\frac{m_{\chi}^{2}}{m_{Z^{\prime}}^{2}}-2\frac{m_{Z^{\prime}}^{2}}{m_{\chi}^{2}}-2 \right]. \label{decay2}
\end{align}
The first term is the decay width of $Z^\prime\to f\bar{f}$ in this model, the second term for $Z^\prime\to \chi_\mu\bar{\chi}_\mu$. For spin-$\frac{3}{2}$ $k_\chi = 1$.

It should be noted that the value of the coupling $g^\prime$ which satisfies the $\Delta a_\mu$ anomaly is $\lesssim 10^{-3}$, where the annihilation cross-section $\sigma(\chi_\mu\bar{\chi}_\mu\to f\bar{f})$ is not large enough for the freeze-out scenario to produce the observed relic density. However, near the resonance $m_\chi \simeq m_{Z^\prime}/2$, there is an enhancement of the cross-section, and the observed relic density may be obtained. The thermally averaged cross section is given as:
\begin{equation}
    \langle\sigma {\rm v}\rangle = \int_{4 m_\chi^2}^\infty \frac{\sigma(s)\sqrt{s-4m_\chi} \,\sqrt{s}\,K_1\left(\frac{\sqrt{s}}{T}\right)}{16\, T \,m_\chi^4\, K_2^2\left(\frac{m_\chi}{T}\right)^2} \,ds,
\end{equation}
where $T$ is the temperature of the thermal bath and $K_{1,2}$ are the modified Bessel functions of order 1 and 2 respectively. This $\langle\sigma {\rm v}\rangle$ can be calculated near the resonance as a function of the ratio $\zeta$ = (DM mass/mediator mass) = $m_\chi/m_{Z^\prime}$ (see for example~\cite{Drees:2021rsg, Dolan:2017osp, Iwamoto:2021fup}).

The expression of relic density is given by
\begin{equation}
    \Omega h^2 = \frac{2.14\times 10^9}{g_\star(x_F)\, M_{\rm{Pl}}\, J(x_F)}.
\end{equation}
Here $M_{\rm Pl}$ is the Planck mass = $1.22 \times 10^{19}$~GeV and
\begin{equation}
    J(x_F) = \int_{x_F}^{\infty} \frac{\langle\sigma {\rm v}\rangle}{x^2} dx, \, \left(x = \frac{m_\chi}{T}\right).
\end{equation}
The decoupling temperature $x_F = \frac{m_\chi}{T_F}$ is obtained by solving
\begin{equation}
    x_F = \ln \left(\frac{0.076 \,M_{\rm {Pl}}\, m_\chi \langle\sigma {\rm v}\rangle}{\sqrt{g_\star} \,\sqrt{x_F}}\right)
\end{equation}
iteratively, where $g_\star$ is the number of relativistic degrees of freedom at freeze-out.
\begin{figure}[t]
  \centering
  \includegraphics[width=0.45\textwidth]{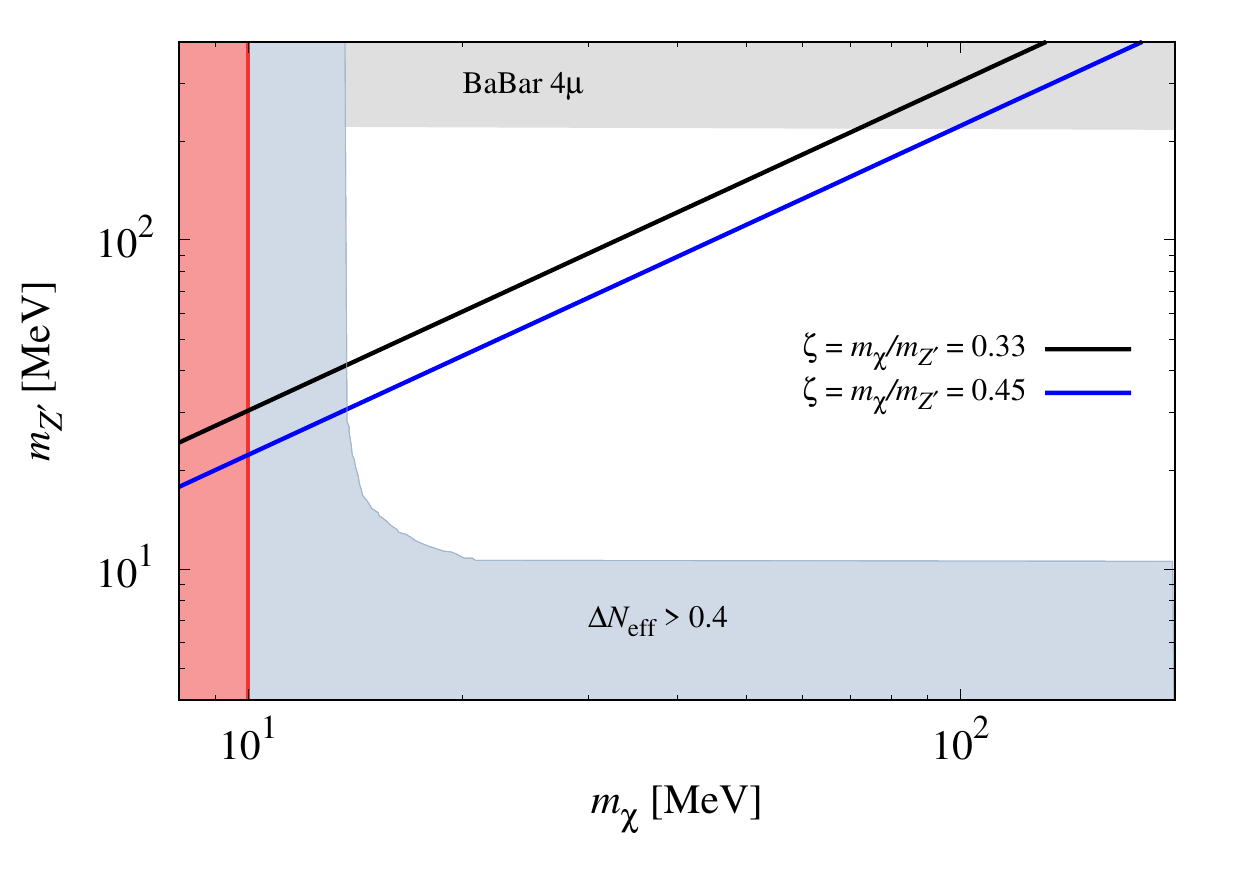}
  \caption{\small The contour plots between the $m_{Z^{\prime}}$ and $m_{\chi}$, for two different ratios $\zeta = m_{\chi}/m_{Z^{\prime}} = 0.33,\, 0.45$ that yield the observed DM density for the couplings $g^{\prime}$ that satisfy $\Delta a_\mu$. The grey, light blue and light red shaded region are excluded by the BaBar experiments, $\Delta N_{\rm eff} > 0.4$ and the CMB and BBN constraints.}
\label{fig:dmamu}
\end{figure} 

In this model the neutrinos acquire an extra energy density through the $Z^\prime_\mu$ decay, as well as from the $\chi_\mu$ annihilation to neutrinos, albeit for $m_\chi \in [1, 100]$~MeV. This can increase the Hubble parameter during BBN, and change the abundance of $He^4$ and $D_2$. In the SM $N_{\rm eff} \sim$ 3.046, whereas from BBN $N_{\rm eff} = 2.878 \pm 0.278$~\cite{Workman:2020zbs,Planck:2018vyg}. We impose $\Delta N_{\rm eff} < 0.4$ at a 95\% confidence level, where $\Delta N_{\rm eff}$ is given in terms of an extra energy density $\delta \rho_\nu$, which contributes to the annihilation of DM and the decay of the gauge bosons $Z^\prime_\mu$ into neutrinos:
\begin{align}
    \Delta N_{\rm eff} = \frac{8}{7}\left( \frac{11}{4}\right)^{\frac{4}{3}}\left. \frac{\delta \rho_\nu}{\rho_\gamma}\right\vert_{T=T_{\nu, D}}, \label{dneff}
\end{align}
where $\rho_\gamma$ is the photon energy density and
\begin{align}
    \delta \rho_\nu = \int \frac{d^3 \vec{p_\chi}}{(2\pi^3)} \frac{g_\chi^\prime E_\chi}{e^{E_\chi/T}+1} + \int \frac{d^3 \vec{p_{Z^\prime}}}{(2\pi^3)} \frac{g_{Z^\prime}^\prime E_{Z^\prime}}{e^{E_{Z^\prime}/T}-1}. \label{drho}
\end{align}
For spin-$\frac{3}{2}$ DM $g_\chi^\prime = 8$ and $g_{Z^\prime}^\prime = 3$ for the $Z^\prime_\mu$ gauge boson. In Fig.~\ref{fig:dmamu} we have shown the contour plots between $m_{Z^\prime}$ and $m_\chi$ for the different ratios $\zeta = m_\chi/m_{Z^\prime} = 0.33$ and 0.45 that yield the observed DM relic density for the coupling $g^\prime$ that satisfies $\Delta a_\mu$. The grey shaded region is excluded by the BaBar experiments~\cite{BaBar:2016sci}. The light blue region is excluded by $\Delta N_{\rm eff} > 0.4$ and the light red region shows the lower bound of 10~MeV for the DM mass, as excluded by the CMB and BBN constraints. 

\section{Summary and Discussion}
\label{summary}
In this work, we first considered a spin-$\frac{3}{2}$ DM particles coupled to neutrinos by making use of an EFT framework with the most general dimension-6 operators. These operators involved scalar, vector and tensor structures (sect.~\ref{eft}), where it was possible to obtain the observed relic density over a wide range of DM mass $\sim 1 - 10^3$~GeV. Note that this was for a cut-off scale $\Lambda \sim 0.5$~TeV to 5~TeV, and where all the couplings were taken to be equal to one. From these considerations, the constraints arising from experimental results~\cite{Super-Kamiokande:2020sgt,Frankiewicz:2015zma,IceCube:2016oqp,Baur:2019jwm,ANTARES:2015vis,DUNE:2015lol} were found to be too weak, $viz.$ $10~{\rm GeV} \lesssim \Lambda \lesssim 800$~GeV, where the EFT framework was not valid for the DM mass range explored and depicted in Fig.~\ref{fig1}. 

In sect.~\ref{nuportal} we then discussed a simple neutrino portal model in which DM coupled to SM neutrinos through a mixing generated by a sterile pseudo-Dirac massive neutrino~\cite{Blennow:2019fhy,Batell:2017cmf}. For this simple scenario there were $s$- and $t$-channel DM annihilation models. Through the exchange of vector mediators, it was found that we could generate the observed relic density for DM masses in the range of 10~MeV to 100~GeV, while the couplings remained in the perturbative regime (Fig.~\ref{fig2}). In the case of the $s$-channel model, the upcoming Hyper-Kamiokande~\cite{Hyper-Kamiokande:2018ofw} experiment provides the best chance of constraining the DM mass in this range.

Finally, in sect.~\ref{u1}, a $U(1)_{L_\mu-L_\tau}$ gauge symmetric model was considered. In this case the gauge boson $Z^\prime_\mu$ was coupled with the same strength to the DM current $J^\mu_{\rm DM}$, and to the $U(1)_{L_\mu-L_\tau}$ current $J^\mu_{L_\mu-L_\tau}$. In exploring the values of these couplings and gauge-boson mass required to satisfy the $\Delta a_\mu$ anomaly (Fig.~\ref{fig:amu}), constraints from the relic density and other cosmological observations were applied. It was observed that the model parameters were severely constrained from several existing and future experiments,  and that much of the parameter space favored by the $\Delta a_\mu$ anomaly could be excluded by the future NA62 and $M^3$ Phase-I experiments. This is highlighted in Fig.~\ref{fig:dmamu} where we were able to show the constraints on the DM and gauge boson masses which would yield the observed relic density and resolve the $\Delta a_\mu$ anomaly. As such, unless the $U(1)_{L_\mu-L_\tau}$ model is ruled out by future experiments it is a potential model which can resolve the $\Delta a_\mu$ anomaly.

\section*{Acknowledgements}
AG thanks SERB, G.O.I. under CRG/2018/004889. MOK was supported by the GES. ASC is partially supported by the National Research Foundation South Africa.



\end{document}